\documentclass[12pt]{aa}
\usepackage[english]{babel}
\usepackage{graphicx}

\onecolumn
\begin{document}

\title{Peculiarities of the atmosphere and envelope of a post--AGB star, the optical counterpart of IRAS\,23304+6347}

\author{V.G. Klochkova, V.E. Panchuk, and N.S. Tavolganskaya}

\institute{Special Astrophysical Observatory RAS, Nizhnij Arkhyz,  369167 Russia}

\date{\today} 

\abstract{Based on high-spectral resolution observations performed with the echelle spectrograph NES 
of the 6--meter telescope, we have studied the peculiarities of the spectrum and the velocity field 
in the atmosphere and envelope of the optical counterpart of the infrared source IRAS\,23304+6347. 
We have concluded about the absence of significant variations in the radial velocity Vr inferred from
atmospheric absorptions and about its coincidence with the systemic velocity deduced from radio data.
The envelope expansion velocity Vexp\,=\,15.5\,km/s has been determined from the positions of rotational band
lines of the C$_2$ Swan~(0;\,0) band. A complex emission--absorption profile of the Swan~(0;\,1) 5635\,\AA{}
has been recorded. Analysis of the multicomponent NaI~D doublet line profile has revealed interstellar
components  V(IS)\,=\,$-61.6$ and $-13.2$\,km/s as well as a circumstellar component with
V(CS)\,=$-41.0$\,km/s whose position corresponds to the velocity inferred from C$_2$ features. 
The presence of the interstellar component with Vr\,=$-61.6$\,km/s in the spectrum allows d\,=\,2.5\,kpc 
to be considered as a lower limit for the distance to the star. A splitting of the profiles for strong 
absorptions of ionized metals (YII, BaII, LaII, SiII) attributable to the presence of a short-wavelength 
component originating in the circumstellar envelope has been detected in the optical spectrum of 
IRAS\,23304+6347 for the first time.}

\authorrunning{\it Klochkova et al.}
\titlerunning{\it Peculiar spectrum  of IRAS\,23304+6347 }

\maketitle

\section{Introduction}

In this paper, we continue to investigate the optical spectra of galactic infrared (IR) sources
identified with highly evolved stars at the short-lived post-asymptotic giant branch (post--AGB) 
evolutionary stage, at which intermediate-mass (3$\div8 \mathcal{M}_{\sun}$) stars rapidly pass into 
the planetary-nebula phase. Our comprehensive studies of supergiants with large IR excesses have led to the 
determination (or reﬁnement) of their evolutionary status. One of the results of our spectroscopy 
for a sample of high luminosity stars, PPN candidates, is the conclusion reached by Klochkova (2012) 
about the inhomogeneity of this sample. Apart from PPNe, the sample produced on the basis of IR photometry 
and low-resolution spectroscopy includes young pre-main sequence stars, high-luminosity stars of various
types, from low-mass semiregular variables to hypergiants.

The fact that PPNe belong to the post--AGB stage makes them extremely interesting both in investigating 
the final evolutionary stages of intermediate-mass stars and in studying the chemical evolution of stars
and galaxies as a whole. The atmospheres of stars at such an advanced evolutionary stage have chemical
peculiarities attributable to the successive change of energy-releasing nuclear reactions accompanied by
a change in the structure and chemical composition of the stellar envelope, the mixing of matter, and the
dredge--up of nuclear--reaction products to the surface layers of the atmosphere. A small homogeneous
subgroup of PPNe with evolutionary overabundances of carbon and heavy metals found in the atmospheres
of their central stars has been identified over the last two decades during the spectroscopy of a sample
of PPN candidates at the world’s largest telescopes  (Klochkova 1995, 2013; Za\v{c}s et al. 1995; Reddy  1997, 
1999, 2002; Klochkova et al. 1999, 2000a, 2000b; van Winckel and Reyniers 2000; Kipper and Klochkova 2006). 
The circumstellar envelopes of these objects have a complex morphology and are generally enriched in carbon, 
which manifests itself in the presence of carbon-containing C$_2$, C$_3$, CN, CO, etc. molecular bands in 
their IR, radio, and optical spectra. These PPNe belong to those few objects whose spectra exhibit the 
21\,$\mu$ envelope emission band (Kwok et al. 1999, Hrivnak et al. 2009). Despite an active search for appropriate 
chemical agents, there is no ultimate identification of this extremely rarely observed feature at present. 
However, its presence in the spectra of PPNe with carbon-enriched envelopes suggests that this emission may
be due to the presence of a complex carbon-based molecule in the envelope (for details and references,
see Hrivnak et al. 2009).

A circumstellar gas--dust envelope manifests itself in peculiarities of the radio, IR, and optical spectra
of post--AGB supergiants. The optical spectra of PPNe differ from those of classical massive supergiants 
by the presence of molecular bands superimposed on the spectrum of an F--G supergiant and
by the anomalous behavior of the profiles for selected spectral features. These can be the complex
emission--absorption profiles of HI, NaI, and HeI lines, the profiles of strong absorptions distorted by
emissions or splitting, and metal emissions. The manifestations of the circumstellar envelope in the optical 
spectra of PPNe are considered in more detail in Klochkova (2014). 

The previous results of our spectroscopy for PPNe with the 6-m BTA telescope were published in a series of original 
papers and are summarized in the reviews by Klochkova (1997, 2012, 2014). In this paper, we present new results 
of high-resolution spectroscopy for the post--AGB star identified with the IR--source IRAS\,23304+6147
(below referred to as IRAS 23304). The central star of IRAS\,23304 is a rather faint (B\,=\,15$\lefteqn{.}^{\rm m}$52, 
V\,=\,13$\lefteqn{.}^{\rm m}$15) supergiant of spectral type G2\,Ia lying  near the Galaxy plane 
(b\,=\,0$\lefteqn{.}^{\rm o}$58). 
According to the high-spatial-resolution Hubble space telescope observations by Sahai et al. (2007), 
the circumstellar envelope in this system has a complex structure including a multipole and an extended halo 
with arcshaped features. 

The first study of the optical spectrum for the central star of IRAS\,23304 belonging  to the group of stars 
with atmospheres enriched in carbon and heavy metals and the calculations of elemental abundances in its 
atmosphere were performed by Klochkova et al. (2000a), who determined the main parameters of the star: 
its effective temperature Teff\,=\,5900\,K, surface gravity, considerably reduced metallicity relative 
to the Sun [Fe/H]\,=\,$-0.61$, and the abundances of 25 other chemical elements. Van Winckel and Reyniers 
(2000) found similar chemical peculiarities based on a higher-resolution spectrum. 
However, in both publications aimed mainly at studying the fundamental parameters of the
star and the chemical composition of its atmosphere, little attention was given to the peculiarities of its
spectrum, the pattern of radial velocities, and their variability with time. 

In this paper, the peculiarities of the optical spectrum for the central star of IRAS\,23304 and their 
variability are considered in more detail. Our observational data are briefly described in
Section~2. The peculiarities of the profiles for the H$\alpha$,  NaI~D lines  metal lines detected from 
high-resolution spectra and molecular bands as well as the data on the velocity field  in the supergiant’s 
atmosphere and envelope are considered in Section~3. Main conclusions are presented in Section~4.

\section{Observational  data}

In this paper, we use the spectra taken at the Nasmyth focus with the NES echelle spectrograph
(Panchuk et al. 2007, 2009) on October 12, 2013. In combination with an image slicer, the NES spectrograph 
provides a spectral resolution R $\approx$60000. A 2048$\times$4096-pixel CCD array has
been used at the NES spectrograph since 2011, which has allowed the recorded spectral range to be
extended considerably. In addition, for comparison, we used the spectra taken with the PFES
echelle spectrograph (Panchuk et al. 1998) at the prime focus of the BTA telescope with a resolution
R$\approx$15000 during several observational sets in 1997.

The details of our spectrophotometric and positional measurements of the spectra were described
in previously published papers; the corresponding references to them are given in Klochkova (2014).
Note that applying the image slicer required a significant modiﬁcation of the standard ECHELLE context
of the MIDAS software package. The data were extracted from two-dimensional echelle spectra with
the software package described by Yushkin and Klochkova (2005). The DECH\,20 code (Galazutdinov 1992), 
which allows, in particular, the radial velocities to be measured from individual features of
complex lines typical for the spectra of the program stars, was used to reduce the extracted spectra.

\section{Peculiarities of the optical spectrum}

\begin{figure}
\includegraphics[angle=0,width=0.6\columnwidth,bb=20 30 570 780,clip]{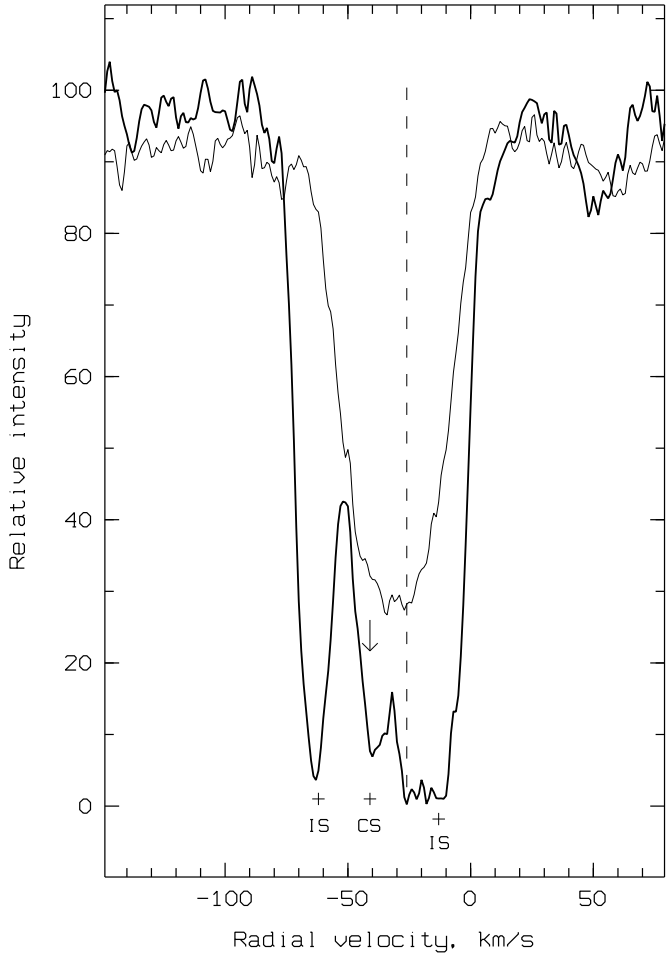}
\caption{\it H$\alpha$ (thin curve) and NaI~D2 (thick curve) line profiles in the spectrum of IRAS\,23304. 
         The arrow marks the envelope velocity inferred from the the C$_2$ Swan band, Vr(CS)\,=$-41$\,km/s. 
         The crosses mark the positions of two interstellar (IS) and circumstellar (CS) components of the NaI~D2 
         line. The vertical dashed line indicates the systemic velocity. Here and in Fig.\,2, the intensity 
         of the normalized continuum along the vertical axis is taken as 100.} 
\end{figure}

It follows from our spectroscopy for the sample of PPNe that the following main types of spectral
features are observed in their optical spectra: (1) low or moderate-intensity metal absorptions whose symmetric
profiles have no apparent distortions; (2) complex neutral hydrogen line profiles changing with time
and including absorption and emission components; (3) the strongest metal absorptions with a low lowerlevel 
excitation potential, their variable profiles are often distorted by envelope features causing an asymmetry 
of the profile or its splitting into components; (4) absorption or emission bands of molecules, mostly
carbon-containing ones; (5) envelope components of the NaI and KI resonance lines; and (6) narrow
permitted or forbidden emission lines of metals originating in the envelopes. The main difference between
the spectra of PPNe and massive supergiants is the presence of features of types 2--6.

All the main peculiarities of the optical spectra for PPNe are contained in the spectra of the post--AGB
star HD\,56126. The latter may be considered by the combination of observed properties (a typical double-humped 
spectral energy distribution, an F--supergiant spectrum with a variable absorption--emission H$\alpha$
line profile, the presence of C$_2$ Swan molecular bands in the optical spectrum originating in an outﬂowing
extended envelope, large overabundances of carbon and heavy metals synthesized during the star evolution 
through the s--process and dredged--up to the surface layers of the atmosphere through mixing) as
a canonical post--AGB star. As follows from Fig.\,2 in the spectral atlas (Klochkova et al. 2007) based on a
long-term monitoring of HD\,56126 with BTA, the H$\alpha$ profile in its spectrum took all of the varieties listed
above: an asymmetric core, a direct or inverse P\,Cyg profile, and a profile with two emissions in the wings.

In the subgroup of PPNe with the 21\,$\mu$ emission feature, the central star of the IR source IRAS\,23304
belongs to the coolest ones (its spectral type is G2\,Ia), with the effective temperature Teff\,=\,5900\,K. The
temperatures in the central stars of IRAS\,22272+5435 (G5\,Ia) and IRAS\,20000+3239 (G8\,Ia) are even
lower: Teff\,=\,5650\,K (Klochkova et al. 2009) and 5000\,K (Klochkova and Kipper 2006), respectively.

\begin{table} 
\bigskip 
\caption{\footnotesize\it Observational data and heliocentric radial velocities Vr, km/s, measured from various spectral features. The number
        of lines used to determine the mean velocity is given in parentheses}     
\medskip 
\begin{tabular}{ l|  c|  c|  c| c| c|  c| c }           
\hline 
 Date&  $\Delta\lambda$, & \multicolumn{6}{c}{\small Vr, km/s}\\      
 \cline{3-8}
     &  \AA{}             & Absorptions   &   H$\beta$&  H$\alpha$ & C$_2$  & NaI    & DIB \\
     &                    &  (297)        &           &            &  (24)  &  (2)   &  (4)  \\  
\hline
12.10.2013  &  4500--6980 & $-25.7\pm0.2$ & $-27.8$   & $-26.6$   &         &$-26.0$ &    \\
            &             &               &           &           &         &$-61.6$ &    \\  
            &             &               &           &           & $-41.3\pm 0.2$ &$-41.0$ &    \\
            &             &               &           &           &         &$-13.2$ & $-14.0\pm1.3$ \\
\hline 
\end{tabular}
\end{table}

{\bf The H$\alpha$ line}.
The H$\alpha$ line in the PPN spectra has complex (a combination of emission and absorption components), time-varying profiles of 
various types: with an asymmetric core, P\,Cyg or inverse P\,Cyg ones, and with two emission components in the wings.
The presence of an emission in the H$\alpha$ line points to a high mass loss rate and is one of the criteria
for searching and identifying PPNe. The H$\alpha$ profile in the spectrum of IRAS\,23304 is also subjected
to significant changes: in Fig.\,1 from Klochkova et al. (2000a), a strong emission is superimposed on
the short-wavelength wing of the absorption profile typical of G--supergiants. The profile in the spectrum
taken in 2013 and presented in Fig.\,1 in ``relative intensity--Vr'' coordinates has an absorption core of
the same depth as that in the 1997 spectrum, but it contains no peculiarities.

{\bf Molecular features}.
In their paper devoted to investigating the molecular component of the PPN spectra, Bakker et al. (1997)
point out the presence of C$_2$ bands for the source IRAS\,23304 whose positions in the spectrum correspond 
to the circumstellar envelope velocity Vr(CS)\,=$-39.7$\,km/s, which leads to the envelope expansion
velocity Vexp\,=\,13.9\,km/s. The observations of two envelope CO bands give the envelope expansion velocity 
Vexp\,=\,9.2 and 10.3\,km/s (Hrivnak et al. 2005). We emphasize that Bakker et al. (1997) pointed out the presence 
of Swan bands in the form of absorption features. The (0;\,0) 5165\,\AA{} band was also recorded in our 
1997 and 2013 spectra. However, the intensity in the head of the (0;\,1) band in both spectra exceeds appreciably the
5635\,\AA{} local continuum level, i.e., we detected a complex  absorption--emission profile of the band (0;\,1) 
5635\,\AA{}.

The table gives the radial velocity Vr(C$_2$)\,=$-41.3\pm0.2$\,km/s that we measured from the positions 
of the 24 rotational lines of the Swan 5165\,\AA{}  band in the 2013 spectrum. Taking into account 
the systemic velocity Vsys(CO)\,=$-25.8$\,km/s from Woodsworth et al. (1990), we obtain the envelope expansion
velocity Vexp\,=\,15.5\,km/s. A velocity Vr(C$_2)\approx -50$\,km/s was derived in Klochkova et al. (2000a)
from the positions of the band heads because of the lower spectral resolution. When an asymmetric head with 
a shade of violet is convolved with a low-resolution, the position of the head is shifted to a short-wavelength velocity,
which explains the difference in the results of the measurements in the 1997 and 2013 spectra.

{\bf The NaI doublet resonance lines and diffuse bands}.
The heliocentric radial velocities for the main components of the NaI~D lines presented in Fig.\,1 are Vr\,=$-61.6$, 
$-41.0$, $-26.0$, and $-13.2$\,km/s  (see the table). Here, it should be noted that the velocities for these components 
differ from those in our  previous publication (Klochkova et al. 2000a). The moderate resolution (R\,=\,15000) in the 
spectra used in this paper was  insufficient for the identification of individual components in the complex NaI~D 
line profile. 

The component of the NaI doublet  lines whose position corresponds to the velocity Vr\,=$-26.0$\,km/s 
originates in the stellar atmosphere, because its  position agrees with the positions of the overwhelming majority 
of symmetric stellar absorptions in the spectrum of  the optical counterpart of IRAS\,23304. 
The longer-wavelength component,  Vr\,=$-13.2$\,km/s, is interstellar and  originates in the Local arm of the Galaxy. 
The shortest-wavelength component,  Vr\,=$-61.6$\,km/s, of the NaI doublet  is also interstellar; it originates in 
the interstellar medium of the Perseus arm. The presence of an analogous  interstellar component with 
Vr$\approx -63$\,km/s in the spectra of the B--stars HD\,4841, HD\,4694, and Hiltner~62,  whose positions 
in the Galaxy are close to the longitude of IRAS\,23304, serves as an argument for this. 
The spectra of these stars, members of the Cas\,OB7 association, were studied by Miroshnichenko et al.~(2009). 
The presence of the component with Vr\,=$-61.6$\,km/s allows us to consider the distance to the Cas\,OB7 association 
d\,=\,2.5\,kpc from Cazzolato and Pineault (2003) as a lower limit for the distance to IRAS\,23304.

Regarding the component of the NaI~D lines with Vr\,=$-41.0$\,km/s, it is natural to assume that it
originates in the expanding circumstellar envelope of IRAS\,23304, where the Swan bands are also formed
(the close radial velocity Vr(CS)\,=$-39.7$\,km/s corresponds to their positions). Thus, we obtain an
envelope expansion velocity Vexp$\approx$13\,km/s typical for PPNe (Loup et al. 1993; Klochkova 2014).

According to the results of Miroshnichenko et al. (2009), there are diffuse interstellar bands (DIBs)
with velocities in the range $-11 \div -14$\,km/s in the spectra of the hot stars HD\,4841, HD\,4694,
and Hiltner\,62 mentioned above. It is also natural to expect the presence of DIBs in the spectra
of the optical counterpart of IRAS\,23304. Luna et al.~(2008) provide their measurements of the positions 
of DIBs that have a very large spread of Vr in the spectrum of IRAS\,23304, from $-26$ to +5\,km/s.
We measured the positions of five absorptions that could be identified with the 5797, 6196, 6203, 6207, and 
6613\,\AA{} DIBs. The mean velocity for them is  Vr(DIBs)\,=$-15$\,km/s. If, however, we discard the 6613\,\AA{}
band blended in the spectrum of the supergiant by a strong YII line, then we obtain a mean velocity 
Vr(DIBs)\,=$-14.0\pm 1.3$\,km/s close to Vr\,=$-13.2$\,km/s inferred from the longestwavelength 
component of the NaI~D lines. We see, that for a more definite conclusion about the positions of DIBs 
in the spectrum  of a cool supergiant, it is necessary to have observational data with an ultrahigh spectral
resolution, R$\ge$100000.

{\bf Asymmetry of the profiles for strong absorptions of ionized metals}.
Thanks to their high spectral resolution (R\,= \,60000), the 2013 observations allowed us to detect
one more, previously unknown peculiarity of the optical spectrum for IRAS\,23304, complex (asymmetric
or split) profiles of the strongest metal absorptions. This peculiarity is clearly seen in Fig.\,2, 
where the YII\,5200\,\AA{}, LaII\,6390\,\AA{}, and BaII\,6141, 6496\,\AA{}, SiII\,6347\,\AA{} line profiles 
are presented. The BaII absorptions in the spectrum of IRAS\,23304 are enhanced to such an extent that 
their equivalent widths W$_{\lambda}$ are comparable to those for the neutral hydrogen lines: 
W$_{\lambda}$(6141)\,=\,0.76\,\AA{}, W$_{\lambda}$(H$\alpha$)\,=\,0.84\,\AA{}. 

Let us consider the detected effect in slightly more detail using the selected lines presented in Fig.\,2 in
``relative intensity -- Vr'' coordinates as an example. As can be seen from Fig.\,2 and the data in the table,
the profiles of these lines include a component whose position coincides with the positions of symmetric
absorptions in the spectrum and a short--wavelength component whose position corresponds to the velocity 
inferred from the C$_2$ Swan band. The proposed interpretation of the complex profile is confirmed by our 
comparison of the line profiles in Figs.\,1 and 2. The position of the short--wavelength component also
coincides with the position of the circumstellar component in the Na~D1 profile.
Thus, it can be asserted that, apart from the photospheric component, the complex BaII line profile
contains a component originating in the circumstellar envelope, suggesting an effecient dredge--up of the
heavy metals produced during the preceding evolution of this star into the envelope. 

The separation between the atmospheric and circumstellar line components is about 15\,km/s. All the lines of
heavy-metal ions (BaII, YII, LaII) in which a profile asymmetry was detected are distinguished by a low
lower-level excitation potential, $\chi_{low} < 1$\,eV. As the spectral resolution is reduced, the intensity of the
envelope components will be added to the intensity of the components originating in the atmosphere,
which will lead to an overestimation of the heavy element abundances determined from strong absorptions. 
The abundances determined from low- and moderate-intensity lines will be more realistic.

\begin{figure}
\includegraphics[angle=0,width=0.6\columnwidth,bb=20 30 570 780,clip]{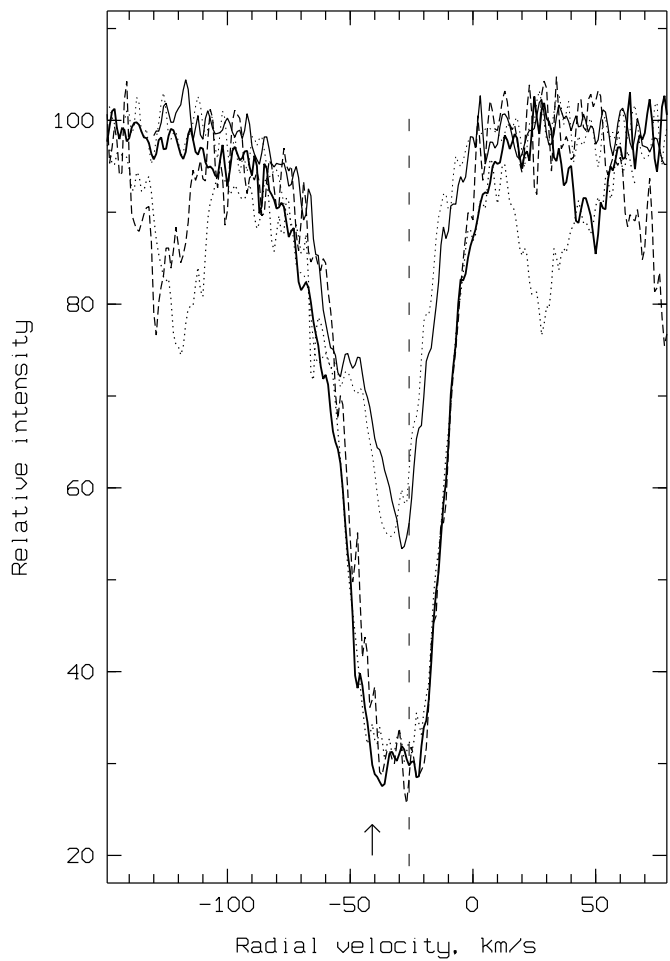}
\caption{\it  Profiles of selected lines in the optical spectrum for the central star of IRAS\,23304. The lower group of lines: 
         BaII\,6141 (thick solid curve), BaII\,6496 (dotted curve), and YII\,5200 (dashed curve). The upper group: LaII\,6390 
         (thin solid curve) and Si\,II 6347 (dotted curve). The arrow marks the envelope velocity Vr(CS)\,=$-41$\,km/s inferred 
         from the C$_2$ Swan band. The vertical dashed line indicates the systemic velocity.} 
\end{figure}

A complex profile for the absorptions of heavy metal ions that, apart from the photospheric component, 
also contains the circumstellar one was found previously in the spectra of the related post--AGB
stars V354\,Lac\,=\,IRAS\,22272+5435 (Klochkova 2009), V448\,Lac\,=\,IRAS\,22223+4327 (Klochkova et al. 2010), 
and V5112\,Sgr\,=\,IRAS 19500$-$1709 (Klochkova 2013). The envelope effect in the spectrum of the high--latitude 
supergiant V5112\,Sgr, which enters the group of PPNe with an atmosphere enriched in carbon and heavy metals and 
its IR spectrum contains the 21\,$\mu$ emission feature, is of greatest interest. An asymmetry and splitting of
strong absorptions with a low lower-level excitation potential were detected in the spectra of V5112\,Sgr
taken with the NES echelle spectrograph at the 6--meter telescope (Klochkova 2013). The effect is maximal for the BaII
lines whose profile is split into three components. The shape of the profiles for the split lines and 
their positions change with time. Our analysis of the velocity field led us to conclude that both short-wavelength
components of the split absorptions originate in the structured circumstellar envelope of V5112\,Sgr.

We emphasize that the strong SiII\,6347 and 6371\,\AA{} lines in the spectrum of the optical counterpart 
of IRAS\,23304 are also asymmetric (Fig.\,2). Apart from the photospheric component, both these
lines include a weak short-wavelength component whose position points to its formation in the stellar
gaseous envelope. This peculiarity of the SiII lines is consistent with a significant silicon overabundance
in the stellar atmosphere (Klochkova et al. 2000a; van Winckel and Reyniers 2000). Thus, for the first
time we have detected the dredge--up of not only s--process elements but also silicon into the envelope.
The synthesis of silicon is possible through the capture of protons by heavier nuclei in the hot layers
of the convective envelope in massive AGB stars with initial masses higher than 4\,${\mathcal M}_{\sun}$. 
A description of this so-called ``hot bottom burning'' (HBB) and the necessary references are available in Ventura
et al. (2011) devoted to the synthesis of Mg, Al, and Si through HBB.

\section{The velocity field in the atmosphere and envelope}

The heliocentric radial velocity of IRAS\,23304 that we measured from a large set of visually symmetric
absorptions  is Vr(abs)\,=$-25.7\pm 0.2$\,km/s. First, this velocity coincides with the systemic velocity 
Vsys(CO)\,=$-25.8$\,km/s inferred from the radio CO observations performed by Woodsworth et al. (1990) 
for a sample of PPNe with the 21\,$\mu$ band. Second, the velocity we found agrees well with the velocity 
estimated from absorptions for two times of observations of IRAS\,23304 in 1994 (Vr\,=$-26$, $-26$\,km/s 
from the data of van Winckel and Reyniers (2000)) and three times of its observations in 1997 
(Vr\,=$-23.4$, $-24.9$, $-25.3$\,km/s from the measurements by Klochkova et al. (2000a)). 

This constancy of the velocity leads us to the preliminary conclusion about the absence of pulsations 
and signatures of binarity in the system. However, it should be noted that Hrivnak et al. (2010) revealed 
a weak brightness variability of the object with an amplitude of about  0$\lefteqn{.}^{\rm m}$2   
and a period P\,$\approx$85\,days as a result of their long--term photometric  monitoring of IRAS\,23304. 
The parameters of the brightness and color variability in IRAS\,23304 derived  by Hrivnak et al. (2010) 
are typical for PPNe with a temperature close to that of IRAS\,23304.  As regards the Vr variability, 
the weak velocity variability from the available, so far scarce data  distinguishes this object among 
the remaining PPNe with enriched atmospheres. Having studied the variability for seven such PPNe, 
Hrivnak et al. (2011) found a pulsational Vr variability with amplitudes of $\approx$10\,km/s.

It is pertinent to dwell on the methodological aspects of studying the velocity field in the atmospheres
of PPNe. The overwhelming majority of the radial velocities used by Hrivnak et al. (2013) were obtained 
with CORAVEL--type spectrometers. For definiteness, consider the results obtained by these authors 
for V354\,Lac (IRAS\,22272+5435). The measurements were performed predominantly by various
cross--correlation techniques. A fundamental shortcoming of the single-channel methods is a mismatch
between the scales of the spectrum and the shifted spectral mask arising at any change in the stellar
radial velocity. In the period 1991--1995, Hrivnak et al. (2013) used the DAO cross--correlation 
photometer (Fletcher et al. 1982; McClure et al. 1985). The above mismatch between the scales was partly
compensated for by a special design of the mask (a with a system of tall slits in the range 4000--4600\,\AA{},
slope changing with distance from the mask center) and a special law of mask motion (at 45$^{\rm o}$ 
to the spectrum axis).

In their description of the DAO correlation photometer, Fletcher et al. (1982) point out a systematic
zero-point drift for the radial velocities (1\,km/s in 2\,h), which they managed to reduce by half, to
0.5\,km/s, by frequent zero-point calibration based on a comparison spectrum. Since the correlation
focus of the DAO 1.2--m photometer at the coude telescope is not subjected to any mechanical deformations, 
the above zero-point drift for the radial velocities is completely determined by the filling of
the spectrograph aperture, variable with the object`s horizontal coordinates.

The Vr measurements in the period 2007--2011 performed from the spectra taken with a CCD array 
in the range 4350--4500\,\AA{} are free from the mismatch of the scales, because the mask is no
longer mechanical but digital; besides, the wavelength range is shorter by four times. The longer
the spectrum interval intercepted by the mask, the more pronounced the mismatch between the mask
and spectrum scales. This problem was solved by T.\,Walraven and J.\,Walraven (1972) by using short
spectral orders, i.e., by applying echelle correlation photometers.

The third series of radial-velocity measurements for IRAS\,22272+5435 was performed with an echelle
correlation photometer whose design, in fact, copies the photometer of Tokovinin (1987). Upgren et al.
(2002) provided the errors calculated from the ``cross-correlation dip area-radial velocity error'' relationship
taken from Jasniewicz and Mayor (1988) for this photometer based on 153 individual measurements
performed for 149 stars. Excluding the errors of more than 1.1\,km/s, we obtained a mean error of
0.83\,km/s for 142 stars from Table\,1 in Upgren et al. (2002). For Tokovinin’s correlation photometer, 
the error due to construction flexure reaches 0.6\,km/s, but it can be reduced to 0.1\,km/s
by orienting the slit parallel to the vertical circle (Tokovinin 1987). However, the errors due to 
atmospheric dispersion shifting the centers of the star monochromatic images across the slit increase in
this case. The angular width of the correlation photometer slit is $\approx$1\,arcsec; the centers of the
monochromatic images for the extreme wavelengths of the simultaneously recorded range (4000--6000\,\AA{})
will be separated virtually by the same amount at a zenith distance of 45$^{\rm o}$. The estimate was obtained
from the tables of differential refraction calculated by Filippenko (1982) for the altitude h\,=\,2 km. 
The differential refraction effect for the Moletai Observatory (h\,=\,0.2\,km) is more pronounced.

Let us estimate the radial-velocity error due to inaccurate centering of the star on the slit. 
For an autocollimation photometer, the scales on the entrance slit and on the mask are identical. 
For $\lambda$\,=\,4400\,\AA{}, a shift of the spectrum by 0.0074\,mm corresponds to a Doppler 
shift by 1\,km/s (Tokovinin 1987).
This means that a radial-velocity measurement error of 1\,km/s can be obtained at an error in centering
the star on the slit of 0.01\,arcsec. An optical element ``tangling'' the rays along the slit width was used as
the slit in Tokovinin’s photometer. The slit width in the photometer used by Hrivnak et al. (2013) is
0.11\,mm (Upgren et al. 2002). The standard star and the program star cannot be unambiguously set on the
slit with an accuracy better than 0.1\,arcsec even in the presence of an autoguider.
On the whole, it can be asserted that the instrumental effects of the single-channel correlation
techniques limit the accuracy of Vr measurements at 0.8\,km/s. This value should be kept in mind
when interpreting the peculiarities of the radial velocity curve for IRAS\,22272+5435 provided by
Za\v{c}s et al. (2009). Recall that periodic radial velocity oscillations with a semi-amplitude of 
3.1\,km/s were detected in this paper, with the overall scatter of individual measurements from 
the mean heliocentric velocity reaching $\pm5$\,km/s. Note that we obtain an accuracy of $\approx$0.8\,km/s 
typical for cross-correlation techniques from the spectra when measuring the position of a single 
line (Klochkova et al. 2010). In the spectra of F--supergiants, the accuracy of Vr from an ensemble 
of several hundred symmetric absorptions is an order of magnitude better.

In addition, in the case of a correlation technique of Vr measurements, the possible peculiarities of the
profiles for strong lines attributable to a complex pattern of the velocity field in the atmospheres of these
stars and the influence of the circumstellar envelope on the profiles are disregarded. Based on the spectra
taken with BTA in a wide wavelength range, we found an asymmetry and splitting of the profiles for low 
excitation lines in the spectra of the post--AGB stars V5112\,Sgr (Klochkova 2013), V354\,Lac (Klochkova 2009), 
and V448\,Lac (Klochkova et al. 2010). This is primarily observed in the BaII, YII, and LaII
resonance absorption lines. A temporal variability of the profiles for the above lines was revealed. The
set of peculiarities of these profiles can be explained by a superposition of spectral features: absorptions
originating in the stellar atmosphere and an envelope emission. The anomalies of the profiles and 
their variability can affect significantly the conclusions about the pulsation properties. For instance, 
for V448\,Lac, Klochkova et al. (2010) detected differential line shifts reaching 8\,km/s and a very low 
pulsation amplitude $\Delta$Vr$\approx$1--2\,km/s, while Hrivnak et al. (2013) found an amplitude exceeding 
this value manyfold. 

The large spread of Vr for close times of observations of V354\,Lac and V448\,Lac (Hrivnak et al. 2013)
at a more regular brightness variation in the star can probably be explained by the neglect of both
instrumental and these subtle kinematic effects. The amplitude of the differential shifts in the atmosphere
of the post--AGB star HD\,56126 is even more significant (Klochkova and Chentsov 2007): they reach
15\,km/s for metal lines. Thus, apart from the pulsational variability with time, the pattern of Vr variations 
in the case of PPNe can also be complicated by differential motions in the extended atmospheres of
the program objects. A detailed analysis of Vr based on high spectral- and time-resolution spectra for the
selected, brightest PPNe allows the differences in the behavior of Vr determined from lines with different
degrees of excitation originating at different depths in the stellar atmosphere to be detected.

\section{Conclusions}

Based on  high-spectral-resolution observations performed with the echelle spectrograph of the
6-m telescope, we studied the peculiarities of the spectrum and the details of the velocity field in the
atmosphere and envelope of a faint supergiant, the central star of the IR--source IRAS\,23304+6347. Our
comparison of the radial velocity Vr\,=$-25.7$\,km/s inferred from numerous low- and moderate-intensity
symmetric absorptions with previously published results points to the absence of significant variations
in the velocity and its coincidence with the systemic velocity deduced from radio data. 

Based on our measurements of the positions for 24 rotational lines of the C$_2$ Swan (0;\,0) $\lambda$\,5165\,\AA{}
band, we determined the envelope expansion velocity Vexp\,=\,15.5\,km/s, typical for post-AGB stars. A
complex emission--absorption profile was detected for the Swan (0;\,1) 5635\,\AA{} band.

Our analysis of the multicomponent NaI\,D doublet line profile revealed interstellar components
with velocities V(IS)\,=$-61.6$ and $-13.2$\,km/s as well a circumstellar component with 
V(CS)\,=$-41.0$\,km/s whose position corresponds to the velocity inferred from C$_2$ features. 
The shortestwavelength component (Vr\,=$-61.6$\,km/s ) of the NaI D lines originates in the interstellar 
medium of the Perseus arm. Its presence allows d\,=\,2.5 kpc to be considered as a low limit for the distance to
IRAS\,23304.

Based on four features identified with diffuse interstellar bands (DIBs), we found the mean velocity 
Vr(DIBs)\,=$-14.0\pm 1.3$\,km/s close to Vr\,=$-13.2$\,km/s inferred from the long-wavelength interstellar 
component of the NaI D lines.
An asymmetry of the profiles for strong absorptions of ionized metals (YII, BaII, LaII, SiII)
attributable to the presence of a short-wavelength component originating in the circumstellar 
envelope in these lines has been detected in the optical spectrum of IRAS\,23304 for the first time.
The overabundance of silicon whose synthesis is possible through the hot bottom burning process in
the hot layers of the convective envelope in massive AGB stars suggests that the star being investigated
belongs to stars with initial masses higher than 4$\mathcal{M}_{\sun}$.

\section*{Acknowledgments} 
This work was supported by the Russian Foundation for Basic Research (project no.\,14--02--00291\,a). 
We used the SIMBAD and ADS astronomical databases.

\section*{References}

\begin{itemize}

\item{}  E.J. Bakker, E.F. van Dishoeck, L.B.F.M. Waters, and T. Schoenmaker, Astron. Astrophys. 323, 469 (1997).

\item{}  F. Cazzolato and S. Pineault, Astron. J. 125, 2050 (2003).

\item{}  A.V. Filippenko, Publ. Astron. Soc. Pacif. 94, 715 (1982).

\item{}  J.M. Fletcher, H.C. Harris, R.D. McClure, and C.D. Scarfe, Publ. Astron. Soc. Pacif. 94, 1017 (1982).

\item{}  G.A. Galazutdinov, Preprint Spec. Astrophys. Observ. RAN, No.\,92 (1992).

\item{}  B.J. Hrivnak and J.H. Bieging, Astrophys. J. 624, 331 (2005).

\item{}  B.J. Hrivnak, K. Volk, and S. Kwok, Astrophys. J. 694, 1147 (2009).

\item{}  B.J. Hrivnak, W. Lu, R.E. Maupin, and B.D. Spitzbart, Astrophys. J. 709, 1042 (2010).

\item{}  B.J. Hrivnak, W. Lu, J. Sperauskas, H. van Winckel, D. Bohlender, and L. Za\v{c}s, Astrophys. J. 766, 116 (2013).

\item{}  G. Jasniewicz and M. Mayor, Astron. Astrophys. 203, 329 (1988).

\item{} V.G. Klochkova, Mon. Not. R. Astron. Soc. 272, 710 (1995).

\item{}  V.G. Klochkova, Astrophys. Bull. 44, 5 (1997).

\item{}  V.G. Klochkova, Astron. Lett. 35, 457 (2009).

\item{}  V.G. Klochkova, Astrophys. Bull. 67, 385 (2012).

\item{}  V.G. Klochkova, Astron. Lett. 39, 765 (2013).

\item{}  V.G. Klochkova, Astrophys. Bull. 69, 279 (2014).

\item{} V.G. Klochkova and E.L. Chentsov, Astron. Rep. 51, 994 (2007).

\item{}  V.G. Klochkova and T. Kipper, Baltic Astron. 15, 395 (2006).

\item{}  V.G. Klochkova, R. Szczerba, V.E. Panchuk, and K. Volk, Astron. Astrophys. 345, 905 (1999).

\item{} V.G. Klochkova, R. Szczerba, and V.E. Panchuk, Astron. Lett. 26, 88 (2000a).

\item{}  V.G. Klochkova, R. Szczerba, and V.E. Panchuk, Astron. Lett. 26, 439 (2000b).

\item{}  V.G. Klochkova, E.L. Chentsov, N.S. Tavolganskaya, and M. V. Shapovalov, Astrophys. Bull. 62, 162 (2007).

\item{}  V.G. Klochkova, V.E. Panchuk, and N.S. Tavolganskaya, Astrophys. Bull. 64, 155 (2009).

\item{}  V.G. Klochkova, V.E. Panchuk, and N.S. Tavolganskaya, Astron. Rep. 54, 234 (2010).

\item{}  S. Kwok, K. Volk, and B.J. Hrivnak, IAU Symp., №191, 297 (1999).

\item{}  C. Loup, T. Forveille, A. Omont, and J.F. Paul, Astron. Astrophys. Suppl. Ser. 99, 291 (1993).

\item{}  R.D. McClure, J.M. Fletcher, W.A. Grundman, and E.H. Richardson, IAU Coll., №88, 49 (1985).

\item{}  A.S. Miroshnichenko, E.L. Chentsov, V.G. Klochkova, S.V. Zharikov, K.N. Grankin, A.V. Kusakin, T.L. Gandet, G. Klingenberg, et al., Astrophys. J., 209 (2009).

\item{}  V. Panchuk, V. Klochkova, M. Yushkin, and I. Najdenov, in Proceedings of the Joint Discussion No.\,4 during the IAU General Assembly of 2006, Ed. by
              I. Gomez de Castro and M.A. Barstow (Editorial Complutense, Madrid, 2007), p. 179.

\item{}  V.E. Panchuk, V.G. Klochkova, M.V. Yushkin, and I.D. Naidenov, J. Opt. Technol. 76, 87 (2009).

\item{}  B.E. Reddy, M. Parthasarathy, G. Gonzalez, and E.J. Bakker, Astron. Astrophys. 328, 331 (1997).

\item{}  B.E. Reddy, E.J. Bakker, and B.J. Hrivnak, Astrophys. J. 524, 831 (1999).

\item{}  B.E. Reddy, D. L. Lambert, G. Gonzalez, and D. Yong, Astrophys. J. 564, 482 (2002).

\item{}  R. Sahai, M. Morris, C. Sanchez Contreras, and M. Claussen, Astron. J. 134, 2200 (2007).

\item{}  A.A. Tokovinin, Sov. Astron. 31, 98 (1987).

\item{} A.R. Upgren, J. Sperauskas, and R.P. Boyle, Baltic Astron. 11, 91 (2002).

\item{}  P. Ventura, R. Carini, and F. D.D’Antona, Mon. Not. R. Astron. Soc. 415, 3865 (2011).

\item{}  T. Walraven and J.H. Walraven, Auxiliary Instrumentation for Large Telescopes, Ed. by S. Lautsen and A. Reiz (ESO, 1972), p. 175.

\item{}  H. van Winckel and M. Reyniers, Astron. Astrophys. 354, 135 (2000).

\item{}  A.W. Woodsworth, S. Kwok, and S.J. Chan, Astron. Astrophys. 228, 503 (1990).

\item{}  M.V. Yushkin and V.G. Klochkova, Preprint Spec. Astrophys. Observ. RAN, No.\,206, (2005).

\item{}  L. Za\v{c}s, V.G. Klochkova, and V.E. Panchuk, Mon. Not. R. Astron. Soc. 275, 764 (1995).

\item{}  L. Za\v{c}s, J. Sperauskas, F.A. Musaev, O. Smirnova, T.C. Yang, W.P. Chen, and M. Schmidt, Astrophys. J. 695, L203 (2009).

\end{itemize}

\end{document}